# Latent unified smooth Hamiltonians for excited state chemistry

David Juergens[1,2], Martin Stöhr[1,2], Andreas E. Hillers-Bendtsen[1,2],

O. Jonathan Fajen[1,2], Todd J. Martínez[1,2*]

[1]Department of Chemistry and The PULSE Institute, Stanford University, Stanford 4305, California, USA

[2]SLAC National Accelerator Laboratory, Menlo Park, 94025, California, USA

[*]Corresponding author: Todd J. Martínez (todd.martinez@stanford.edu)

## Abstract

We describe a neural network architecture and training procedure designed to model electronic ground and excited states of arbitrary molecular systems. By indirectly learning a latent, implicit basis representation of the electronic-state Hamiltonian, the model offers a unified treatment of multiple electronic states, conical intersections, and non-adiabatic couplings. The formalism can be further extended to learn consistent latent representations of additional operators such as transition dipole moments, for example. To demonstrate the general capabilities of our architecture, we train and evaluate networks on two realistic photochemical systems, thymine and azobenzene. The resulting models accurately reproduce energies and oscillator strengths for the ground- and low-lying excited states relevant to the photochemistry of these systems. We highlight the performance of the trained networks by studying critical molecular geometries, including conical intersections and excited state minima. By construction, the proposed framework also recovers the emergence of Berry phase accumulation around conical intersections. By pairing key mathematical structure from quantum chemistry with the representation learning power of transformers, the presented architecture offers a qualitatively new path toward fast and accurate ground- and excited-state simulations.

## Introduction

Photochemical processes are at the heart of fundamental biological phenomena such as vision and photosynthesis as well as sustainable energy solutions from artificial light harvesting to photocatalysts.[1-6] Computational modeling can offer detailed atomistic insights for a better understanding of these processes, eventually enabling targeted optimization for technological and biological applications.[7-9] The reliable *in silico* description of electronically excited states and photochemistry requires a highly sophisticated treatment of the electronic structure, often only provided by advanced *ab initio* methods such as second-order perturbation-corrected complete active space self-consistent field calculations.[10-13] While providing the required level of accuracy, these methods come at dramatic computational costs, which considerably limits their application to realistic problems in excited state chemistry. The first reason for this is that photochemical processes are highly non-equilibrium, often requiring extensive dynamics simulations together with sufficient sampling of initial conditions. Second, excited state chemistry can be highly sensitive to the molecular environment.[14-19] Evolution, for example, has employed this by mutating residues in proteins to tune photoactive sites from afar.[20] As a result, the time and length scales involved in many biologically and technologically relevant photochemistry applications quickly become intractable for the available high-level quantum chemistry methods.

In the context of ground-state chemistry, the introduction of machine learning interatomic potentials (MLIPs) has transformed the field of molecular simulations in recent years. Promising nearly *ab initio* accuracy at orders-of-magnitude reduced computational cost, MLIPs are on the way to turning large-scale computational modeling of molecules and materials into routine tasks.[21-28] As of now, this success has not broadly translated to excited state simulations. This is because — contrary to ground state potential energy surfaces — the excited energy landscape is a set of tightly coupled functions resolving multiple state energies at each molecular geometry. As a naive approach, one can build multi-state MLIPs that emit multiple scalar energies instead of just one.[29-33] The fundamental downside of this approach is that adiabatic electronic eigenstates are non-smooth when intersecting, showing a conical character.[34-39] To address smoothness requirements away from conical intersections, modern MLIPs, on the other hand, are deliberately designed to be smooth.[21,40-44] As a result, a naive multi-state MLIP approach fails to accurately describe state intersections.[12,44,45] Given the critical role of conical intersections in excited state chemistry,[14,15,34-37,46] this leads to considerable limitations for current MLIPs in photochemistry simulations. Training an MLIP on surfaces it is fundamentally unable to reproduce inadvertently leads to problematic artifacts such as spurious avoided crossings or double-crossings of artificially smoothened cones where the "excited state" falls below the supposed ground state.[44,47]

A promising alternative is to follow the formalisms in quantum chemistry, where the electronic energies are obtained as the eigenvalues of a Hamiltonian matrix parametrized by the nuclear coordinates, $\mathbf{H}(\mathbf{R})$.[12,48-52] In principle, the full $N$-dimensional system Hamiltonian includes an infinite number of states, making this representation impractical. However, the vast majority of photochemistry involves only a small number of low-lying electronic states. This motivates the use of a finite submatrix, $\mathbf{H}(\mathbf{R}) \in \mathbb{R}^{m \times m}$ with $m \ll N$. In quantum chemistry, this $m$-state electronic Hamiltonian matrix is most commonly represented in its "adiabatic" (*i.e.*, diagonal) form, but it can also be expressed in a "diabatic" form which minimizes the derivative coupling between state basis functions. In the context of machine learning, both representations have important limitations. Due to the non-Abelian Berry curvature of electronic energies, there exists no unique or global diabatic representation for polyatomic molecules.[53] Preparation of smooth reference data across molecular configurations by combining locally-diabatic representations can be cumbersome and technically demanding, further complicated by the Berry phase when including conical intersections. Although learning an explicit diabatic form can be a viable option for the local description of electronic states in some cases,[48-51] it quickly becomes impractical for more transferable models or when many different electronic states are involved. Adiabatic reference energies are ordered solutions to the electronic eigenvalue problem in a uniquely defined representation, which generalizes more easily across configuration space. However, the energetic ordering leads to derivative discontinuities in the target energies around conical intersections, which cannot be resolved by smooth neural network potentials. This issue has previously been approached by learning the geometry-dependent characteristic polynomial of a potential matrix and diagonalizing the corresponding companion matrix.[47,54,55] Following a related route, the X-MACE model[45] encodes adiabatic reference energies using a DeepSet embedding and trains an MLIP to predict the encoded permutationally-invariant representation. For inference, predicted DeepSet representations are again decoded into ordered energy labels by constructing and diagonalizing a Hermitian companion matrix. While representing a major step towards a neural network-based description of excited state manifolds, these approaches do not provide a unified description of energetics and other important properties such as transition dipole matrix elements and nonadiabatic couplings (NACs).

To address the remaining challenges, we propose the Latent Unified Smooth Hamiltonian (LUSH) approach. In contrast to previous efforts to produce adiabatic or diabatic representations of the electronic-state Hamiltonian, LUSH is a neural network aiming to learn a general (*i.e.*, neither strictly diabatic nor adiabatic) latent Hamiltonian in an implicit basis representation. This avoids the aforementioned limitations of diabatization and allows us to smoothly interpolate the constructed operator across configuration space. Diagonalization of this Hamiltonian yields adiabatic energy levels including conical intersections. Furthermore, NACs are readily computed by taking its derivatives and employing the Hellman-Feynman relation. Finally, the adiabatic eigenvectors allow formulation of a unified framework to evaluate general electronic (transition) properties. Specifically, we can generate latent representations of operators beyond the Hamiltonian and rotate their implicit basis form into the adiabatic frame.

## A neural network for electronic-state Hamiltonians

In the following, we summarize the LUSH framework and neural network architecture, which predicts adiabatic state energies and gradients *via* the generation of a general electronic-state Hamiltonian matrix. We further outline the calculation of electronic (transition) properties, such as NAC vectors and transition dipole moments, all in a consistent latent-to-adiabatic transformation. We conclude this section with a description of the training procedure and loss definition.

### Network architecture

We begin with the network input, recalling the aim of generating a set of fixed-size operator representations, $\mathbf{O}(\mathbf{R})$, one of which must be the Hamiltonian matrix, $\mathbf{H}(\mathbf{R}) \in \mathbb{R}^{m\times m}$, which is diagonalized to obtain adiabatic energies. Molecules of arbitrary size are first featurized as a graph, with node features derived from one-hot atom types and edge features derived from radial basis function embeddings of interatomic distances. An SE(3)-invariant message passing neural network adapted from Ref. [56] embeds atom-local geometric information into $n$ per-atom features, each of dimension $d$. Critically, we now require an operation which converts this arbitrary-size array into a fixed-size array of $m$ tokens ($d$-dimensional feature vectors) while maintaining a rich description of the molecular geometry. To do this, we use a cross attention update with a set of $m$ trainable model latent vectors.[57,58] This projects the entire set of nodes into a length- $m$ array of vectors $\mathbf{S} \in \mathbb{R}^{m\times d}$, which we denote as the single-state representation. After creation, $\mathbf{S}$ is readily refined using a standard self attention stack to yield $\mathbf{S}' \in \mathbb{R}^{m\times d}$ and conclude the geometry encoding step (Figure 1A).[59]

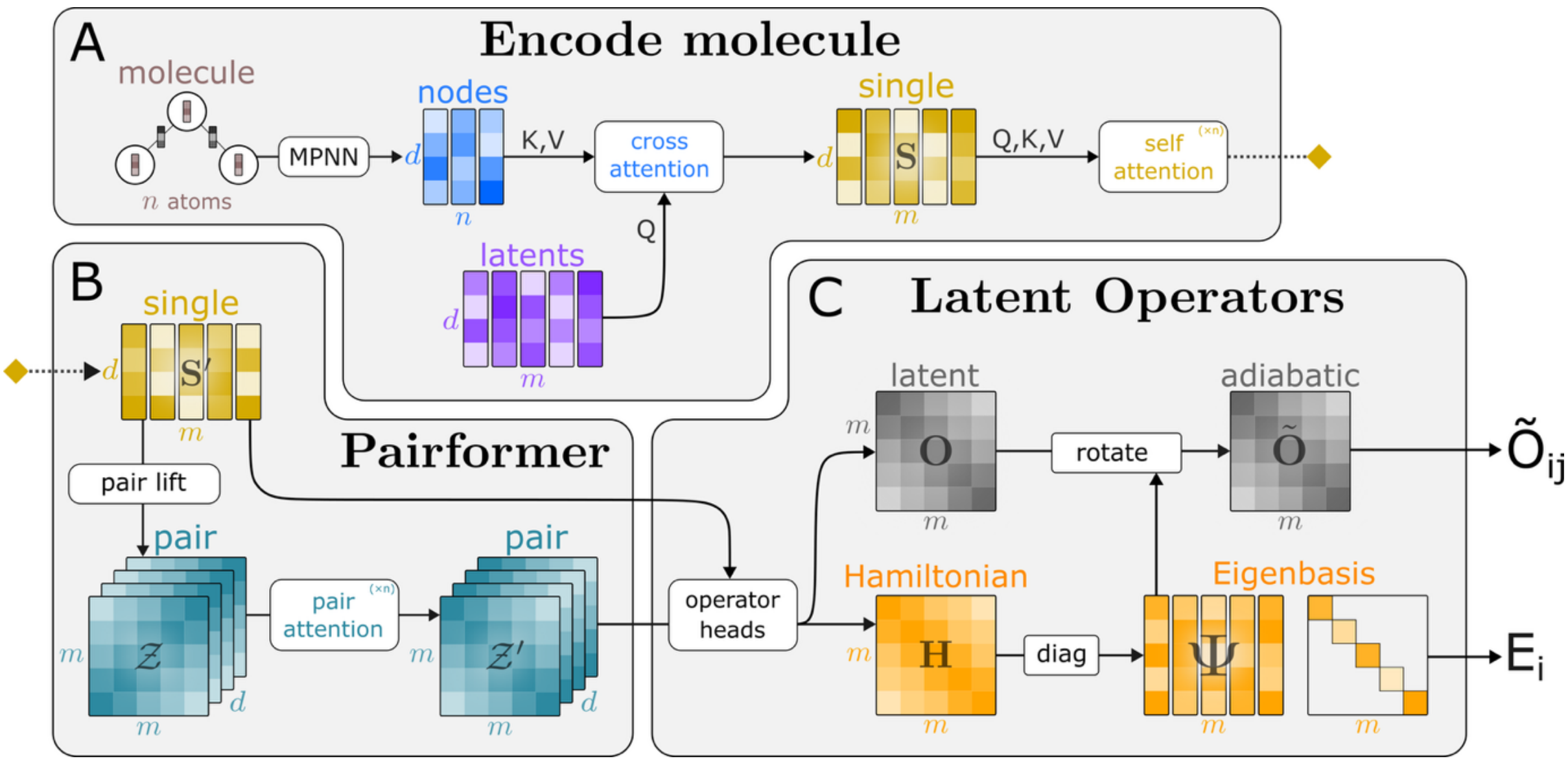

**Figure 1: Overview of the LUSH architecture** (A) Arbitrary molecules first undergo a geometry encoding step, where a message passing neural network (MPNN) encodes atom-local geometric information into node features on a graph (blue "nodes"). A cross attention update with trainable model latents (purple) maps the variable-length ($n$) set of node features into a fixed-length ($m$) array of tokens called the initial "single-state" representation ($\boldsymbol{S}$, yellow). $\boldsymbol{S}$ is then refined through a series of self attention updates yielding the refined single-state representation, $\boldsymbol{S}'$. (B) An initial state-pair representation $\mathcal{Z}$ of the single-state tokens is constructed through a "pair lift" on $\boldsymbol{S}'$, in which all $(\boldsymbol{S}_i', \boldsymbol{S}_j')$ are combined in a pairwise representation. A series of pairwise attention updates then refines $\mathcal{Z}$ to obtain $\mathcal{Z}'$. (C) Parallel network heads map $\mathcal{Z}' \in \mathbb{R}^{m\times m\times d}$ to a set of latent operator representations $\boldsymbol{O}$, one of which is $\boldsymbol{H}$, the latent Hamiltonian. $\boldsymbol{H}$ is diagonalized to yield electronic state energies $E_i$ and a set of state-associated eigenvectors $\boldsymbol{\Psi}$. All other latent operators are rotated into the eigenbasis $\boldsymbol{\Psi}$ to yield final (adiabatic) operator representations.

From the refined single-state representation, we aim to build a representation that can ultimately be consumed by simple, parallel network heads to produce the latent operator representations. In contrast to predicting individual scalars and arranging them into a matrix, LUSH aims to explicitly encode the matrix structure of $m \times m$ operator representations in its architecture. To achieve this, we first "pair-lift" $\mathbf{S}'$ into an intermediate representation, $\mathcal{R} \in \mathbb{R}^{m\times m\times 3d}$ through the following operations

$$\mathcal{R}_{ijk} = \begin{cases} \mathbf{S}'_{ik} \cdot \mathbf{S}'_{jk}, & \text{if } k < d \quad \text{(Hadamard product)} \\ \mathbf{S}'_{ik} \cdot \mathbf{S}'_{jk}, & \text{if } d \le k < 2d \quad \text{("Hadamard sum")} \\ \left|\mathbf{S}'_{ik} - \mathbf{S}'_{jk}\right|, & \text{if } 2d \le k < 3d \end{cases} \tag{1}$$

Subsequently, $\mathcal{R}$ is projected into an initial state-pair representation, $\mathcal{Z} \in \mathbb{R}^{m\times m\times d}$, by a single-layer perceptron followed by a resnet block. The elements of a meaningful representation of $\mathbf{H}$ naturally show complex interdependencies because they correspond to a common basis of a single operator and due to the additional structure imposed by the underlying many-body electronic-structure problem. To reflect these interdependencies in LUSH, we next incorporate global communication among all $\mathcal{Z}_{ij}$. By working with a state-pair matrix/tensor representation rather than an array of individual scalars, this is accounted for more directly. All-to-all coupling can be achieved with a number of available image or pairwise attention architectures.[60-62] We select the Pairformer update from AlphaFold3, since its explicit inclusion of interactions among three $\mathcal{Z}_{ij}$ vectors should facilitate the required global communication[63] (for further details on the LUSH Pairformer, see Methods). After refinement, the resulting $\mathcal{Z}'$ contains the final state-pair representation (Figure 1B).

**Electronic (transition) properties from unified operator representations**

Figure 1C displays the scheme for operator generation from the refined pair representation. In parallel, a set of operator heads (multilayer perceptrons or resnets) each map $\mathcal{Z}' \in \mathbb{R}^{m\times m\times d}$ to latent matrix/tensor representations of the respective operator, $\mathbf{O}(\mathbf{R})$. Among these is the Hamiltonian $\mathbf{H}(\mathbf{R})$, which is symmetrized and diagonalized to produce state energy predictions $E_i$ and a corresponding set of eigenvectors $\Psi_i$, representing adiabatic pseudo-wavefunctions. To ensure stability through differentiation of eigenvalue problems with degenerate eigenstates, diagonalization is performed using a custom PyTorch implementation as presented in Refs. [64] and [65]. By rotating all other operator matrices/tensors from their initial representation into this adiabatic basis, a representation of every operator in a consistent implicit basis is favored in the loss landscape.

It is worth examining the interpretation of the operator representations $\mathbf{O}(\mathbf{R})$ before and after their transformation into the adiabatic basis $\mathbf{\Psi}$. A matrix representation of an operator $\hat{O}$ is simply a recording of how $\hat{O}$ acts on a chosen set of (basis) vectors. Specifically, given a set of basis states $\{|p\rangle\}$, the matrix element $\mathbf{O}_{pq} = \langle p|\hat{O}|q\rangle$. The initial operator representations generated from $\mathcal{Z}$ thereby correspond to a general, implicit electronic-state basis. Diagonalization of the respective Hamiltonian, $\mathbf{H}$, then produces the adiabatic basis. In order to obtain a unique representation of electronic (transition) properties, we rotate the initial general operator representations into the adiabatic frame of the latent Hamiltonian. We denote these transformed operator representations as $\tilde{\mathbf{O}}$. With this, the diagonal elements $\tilde{\mathbf{O}}_{ii} = \langle \Psi_i | \hat{O} | \Psi_i \rangle$ correspond to the expectation value of an operator on a particular adiabatic electronic state $\Psi_i$. The off-diagonal elements $\tilde{\mathbf{O}}_{ij}$ refer to "transition" properties between states $i$ and $j$. As an example, in the current work we produce transition dipole moment tensors $\tilde{\mu} = \mathbf{\Psi}^{\dagger} \mu \mathbf{\Psi}$ with $\tilde{\mu}_{ij}$ representing state-resolved molecular dipole moments on the diagonal and transition dipole moments between states on the off-diagonal.

The procedure of generating electronic state and transition properties by first producing latent operators in an implicit basis representation and subsequently rotating them into the adiabatic frame has three important implications for network training and the calculation of key excited state properties. First, diagonalizing $\mathbf{H}$ relieves the network of the task of learning to produce the required derivative discontinuities at conical intersections. The network instead learns smooth $\mathbf{H}_{pq} = \langle p | \hat{H} | q \rangle$ elements in the general electronic-state basis. Second, the eigenvalues and -vectors $E_i$ and $\Psi_i$ derived from $\mathbf{H}$ enable the direct calculation of NAC vectors $\mathbf{d}_{ij} = \langle \Psi_i \left| \frac{d}{d\mathbf{R}} \right| \Psi_j \rangle = \left\langle \Psi_i \left| \frac{d\mathbf{H}}{d\mathbf{R}} \right| \Psi_j \right\rangle / (E_j - E_i)$ between states. This formulation is known to be exact when the space spanned by the chosen electronic states is geometry independent (and to be very accurate when the geometry dependence is weak). Directly calculating NACs in strict agreement with the eigenvalue problem of a symmetric, parametric matrix (the general electronic-state Hamiltonian, $\mathbf{H}(\mathbf{R})$) has the additional benefit of correctly reproducing the topographic features of the NAC vectors. This naturally encodes branching planes, the plane spanned by the gradient of the state energy gap and the derivative coupling, in which the energy degeneracy between states is lifted. In addition, LUSH is by definition guaranteed to correctly have the conical intersection as a singularity in the NAC line field (NAC vectors form a vortex-like field around the intersection, as discussed for Figure 2 below). From a mathematical perspective, this topographical feature is characteristic for the eigenvalue problem of symmetric matrices with parametric dependence and is therefore by no means guaranteed when instead predicting NACs *via* a separate network head as in previous architectures.[31-33,45] From a quantum mechanical perspective, the vortex-like line field is responsible for the accumulation of the Berry/geometric phase when circling a conical intersection.[66-68] Finally, inclusion of additional latent operators and their explicit rotation into the adiabatic frame should improve refinement of $\mathcal{Z}$ by imposing a constraint toward a unified implicit basis for all operators. Thus, while still offering the scaling and representation learning advantages of transformer-based neural networks, the LUSH architecture inherently encodes crucial fundamental features of electronic state energies, their intersections, and coupling.

**Training procedure and loss definition**

To train the network, we can begin with a dataset $\{(X, Y)\}$ where each data point $(X, Y)$ contains the network inputs $X$ and target values $Y$. In the current work the inputs $X$ are the atomic numbers $\mathbf{Z}$ and cartesian coordinates $\mathbf{R}$ of nuclei in a molecular system. Additional inputs such as charge or spin multiplicity are readily added if desired but not used here as all studied systems are neutral, closed-shell species. The targets $Y$ are electronic (transition) properties, or quantities derived from them, which we would like the network to predict given the input $X$.

Care must be taken to only select properties $Y$, or apply loss functions to them, that are invariant to the phase of the eigenvectors.[45,69,70] In the present study there are three targets: the *ab initio* electronic state energies relative to the sum of ground state energies of the isolated atoms, $E_i$ , the excitation energies $\Delta_{ij} = E_j - E_i$ computed from the predicted state energies, and oscillator strengths $f_{ij} = \frac{2}{3} \|\tilde{\mu}_{ij}\|^2 / \Delta_{ij}$ calculated from the transition dipole, $\tilde{\mu}_{ij}$, and $\Delta_{ij}$, between states $i$ and $j$. Given an input, LUSH generates predictions for the three targets on which we can apply a loss to train the model. In the current study, we use the following loss function:

$$L = L_{\mathrm{E}} + L_{\Delta} + L_{\mathrm{osc}} \tag{2}$$

where $L_{\mathrm{E}}$ is a mean squared error (MSE) calculation applied to the energies $E_i$, $L_{\Delta}$ is a smooth-L1 (*i.e.*, a scaled Huber[71]) loss applied to absolute errors in excitation energies $\Delta_{ij}$, and $L_{\mathrm{osc}}$ is a smooth-L1 applied to absolute errors in predicted oscillator strengths $f_{ij}$. Note that because $L_{\Delta}$ and $L_{\mathrm{osc}}$ contain quantities derived from the state energy predictions $E_i$, gradients from all three loss terms supervise the predicted $E_i$.

An important limitation of the $m$-dimensional Hamiltonian submatrix arises at intersections at the boundary of its spectrum. At an $m/(m+1)$ intersection, the matrix representing the lowest $m$ eigenstates of the full Hamiltonian becomes fundamentally non-smooth and, in fact, discontinuous. In the spectral picture, the lower-lying states remain smoothly interpolatable, but the $m^{\mathrm{th}}$ eigenvalue has a derivative discontinuity and the corresponding eigenvector is discontinuous. Because all states are coupled through the eigendecomposition, naively training a LUSH model on all $m$ states can lead to contamination of *all* obtained states in regions near $m/(m+1)$ intersections. In order to isolate the error in the problematic highest state(s), we introduce an adaptive loss function with state weighting similar to approaches for diabatic interpolation[72] or complete active space self-consistent field simulations.[73,74] During training, the loss contribution of a given state, $i$, is weighted based on the ground truth state energy, $\bar{E}_i$, according to

$$w_i = \begin{cases} \frac{1}{2}\left(1 - \cos\left(\frac{\pi}{\tau}(\bar{E}_m - \bar{E}_i)\right)\right) & \text{if } 0 \leq \bar{E}_m - \bar{E}_i < \tau \\ 1 & \text{else} \end{cases} \tag{3}$$

smoothly switching from one to zero as the state energy comes within $\tau$ of the highest considered state ($\bar{E}_m$). For the present work, we chose $\tau = 0.5$ eV. Properties characterizing transitions between states $i$ and $j$ receive a weight of $w_i \cdot w_j$. In this setup the $m^{\mathrm{th}}$ state is unsupervised, essentially providing a pure buffer dimension for the smooth interpolation of the remaining spectrum. Furthermore, the model is allowed to make mistakes in the $(m-1)^{\mathrm{th}}$ state whenever near an $(m-1)/m$ intersection. The lowest $(m-2)$ states, finally, are fully accounted for. This allows us to isolate the error arising from the fundamental limitation of the truncated Hamiltonian matrix to the $(m-1)^{\mathrm{th}}$ state. Importantly, the $(m-1)^{\mathrm{th}}$ state is also correctly described away from the $(m-1)/m$ intersection, which allows the model to faithfully reproduce the $(m-2)$ lowest-energy states including its intersections (as long as they are sufficiently far from the $(m-1)/m$ intersection seam). So, for every LUSH model, we choose a state-weighted loss and treat (at least) the two highest-lying eigenstates as buffer states. For simplicity, the above discussion is limited to two-state degeneracies, although one can extend this to encompass multistate intersections also.[75-77] To further safeguard against such degeneracies at the spectrum boundary or $(m-2)/(m-1)$ intersections near $(m-1)/m$ degeneracies, one can increase the number of buffer states.

## Results

In the following, we highlight the capabilities of the LUSH architecture by studying its performance on ground-state energetics across the QM9 dataset and testing its ability to reproduce excited state landscapes for select molecules. Specifically, we showcase the description of conical intersections and NACs between the first ($S_1$) and second ($S_2$) excited singlet state of thymine and explore the excited state manifold of azobenzene including scans of critical degrees of freedom, ground- and excited-state minima, and minimum energy conical intersections (MECIs).

### Chemical accuracy for ground state energetics

As a first test of our architecture and training scheme, we evaluated its ability to learn ground state energies on the well-studied benchmark dataset QM9, which contains density functional theory (DFT) calculations at the B3LYP/6-31G(2df,p) level of theory for 134,000 different small organic molecules.[78] Figure S1 displays the results of training a 5.95 million parameter LUSH network on the QM9 total potential energy prediction task. For the published validation set of 100 randomly sampled molecules, the model approaches chemical accuracy (~1 kcal/mol) in about 48 hours on a single NVIDIA Tesla V100 (32GB) GPU. For this experiment, we set the number of model latents (controlling the maximum number of eigenvalues emitted) to $m = 8$, and take the lowest eigenvalue $E_0$ as the predicted total energy while the remaining eigenvalues $E_{1-7}$ are unused and unsupervised. Note that because QM9 is a ground-state DFT dataset, the excited state loss function components $L_\Delta$ and $L_{\text{osc}}$ are not used in this experiment.

### Recovering conical intersections and NAC fields

We next investigated whether LUSH can capture key electronic structure features that govern photochemical dynamics, including excitation energies, oscillator strengths, conical intersections and NACs. We selected thymine as a representative system for testing this because recent experimental and theoretical studies have found that, following photoexcitation (into the bright $S_2$ state), it undergoes rapid radiationless relaxation to $S_1$ through a conical intersection.[79-83] Thymine therefore provides a realistic setting in which the behavior of LUSH near geometries with strong NAC and electronic state degeneracy can be examined. Thymine is one of the nine photoactive species present in the recently published QeMFi dataset, which contains linear response time-dependent density functional theory (TDDFT) calculations with a variety of basis sets on Wigner-sampled geometries of organic molecules in gas phase.[84,85] We trained a 7.01 million parameter LUSH network on the 135,000 TDDFT-CAM-B3LYP[86]/def2-TZVP[87] results in QeMFi (15,000 per molecule). As shown in Figure S2, training converged rapidly toward models with low validation errors for total energies, excitation energies, and oscillator strengths across all nine QeMFi species.

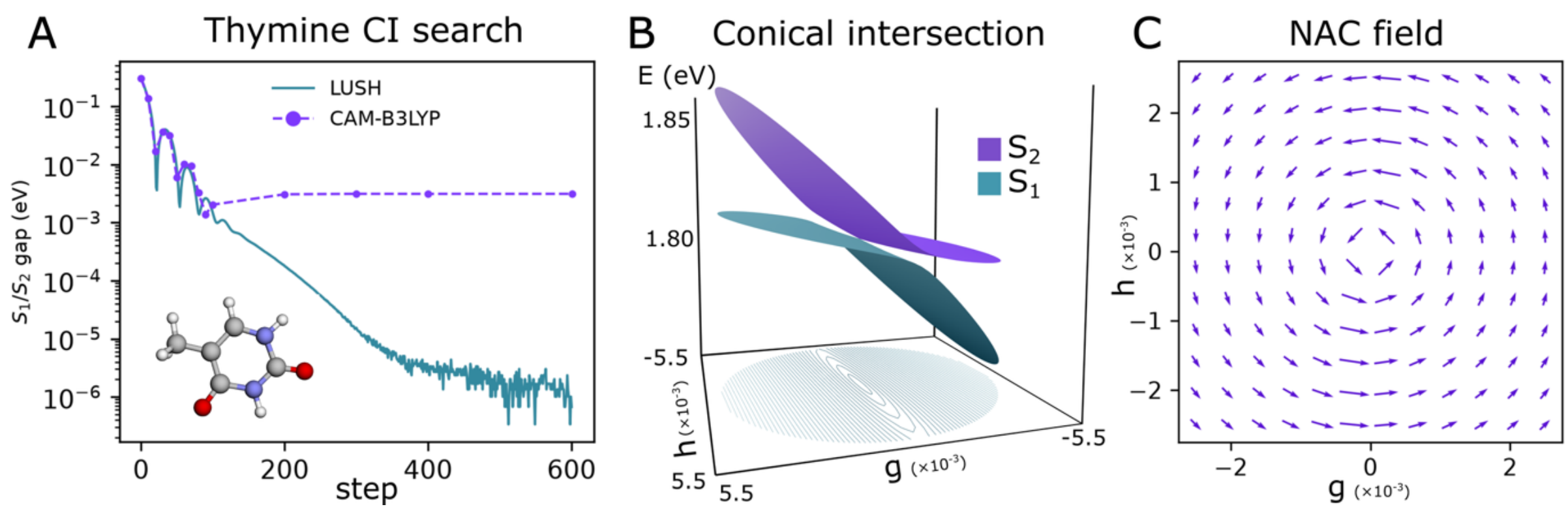

**Figure 2: Thymine conical intersections with LUSH** (A) Convergence of conical intersection (CI) search between $S_1/S_2$ on the LUSH surfaces while also tracking the *ab initio* gap at various points during the optimization. (B) LUSH energy surfaces around the identified CI in the branching plane as defined by the two-dimensional space in which the energy degeneracy is lifted. (C) Quiver plot of the projection of nonadiabatic coupling vectors ($\mathbf{d}_{12}$) onto the branching plane around the CI. For clarity, projections are rescaled by the 0.825th power of their norm.

We first evaluate the model resulting from this training session at critical thymine geometries obtained at the TDDFT level. Starting from the $S_1/S_2$ MECI geometry obtained with equation-of-motion coupled cluster with singles and doubles,[83] we minimized thymine on the $S_1$ TDDFT surface using the TeraChem electronic structure program.[88,89] We then took the geometries along this optimization path and predicted their energies and oscillator strengths with the trained network. Figure S3 compares the network predictions with the *ab initio* values across these held-out geometries and confirms a close agreement for $S_0 \rightarrow S_n$ excitation energies and oscillator strengths. Notable is the ability of the network to precisely track the $S_3$ and $S_4$ states through their avoided crossing, as indicated by a flipping of the $S_0 \rightarrow S_3$ and $S_0 \rightarrow S_4$ oscillator strengths matching TDDFT.

Encouraged by these results, we then searched for thymine conical intersections directly on the network's learned surfaces. To do so, we relied on PyTorch's automatic differentiation[90] framework to obtain the gradient of the state gap and perform a simple Adam[91] optimization to reach the intersection seam. Figure 2 displays the results of performing an $S_1/S_2$ CI search on thymine with the LUSH surfaces. Although the QeMFi dataset contains no $S_1/S_2$ CIs (Figure S4), LUSH does find a thymine geometry with predicted zero gap (Figure 2A). Moreover, this geometry is also near the *ab initio* intersection seam, with a TDDFT $S_1/S_2$ gap of $3.1 \cdot 10^{-3}$ eV – less than half of the smallest gap observed in QeMFi. We can examine the CI structure by computing the vectors $\mathbf{g} = \partial(E_j - E_i)/\partial\mathbf{R}$ and $\mathbf{h} = \left\langle \Psi_i \middle| \frac{d\mathbf{H}}{d\mathbf{R}} \middle| \Psi_j \right\rangle = \mathbf{d}_{ij}(E_j - E_i)$ at the point of degeneracy. The consistent treatment of state energies and NACs in LUSH guarantees that $\mathbf{g}$ and $\mathbf{h}$ define the only two nuclear degrees of freedom which lift the degeneracy around a two-state conical intersection. Plotting the $S_1$ and $S_2$ energy surfaces in the plane spanned by $\mathbf{g}$ and $\mathbf{h}$ (the branching plane), Figure 2B confirms the conical shape of the LUSH intersection.

Beyond geometrically correct conical intersections, accurate photochemical simulations also rely on the consistent computation of the NAC vectors. A fundamental result of the LUSH architecture is that, like an *ab initio* intersection, $\mathbf{d}_{ij}$ are guaranteed to be singular at degeneracies. By design, LUSH also produces the characteristic vortex structure of the NAC field around conical intersections responsible for the emergence of the Berry/geometric phase when the nuclei trace a path encircling the intersection. Figure 2C demonstrates this effect in LUSH by plotting the projections of $\mathbf{d}_{ij}$ onto the branching plane at various locations around the intersection. The NAC projections reflect the topology of the cone near the intersection, with small components in $\mathbf{h}$ corresponding to slow variation of the energy gap along this coordinate. Along $\mathbf{g}$ the energy gap varies more rapidly (Figure 2B) and correspondingly the projections have larger components in the $\mathbf{g}$ direction, particularly around $\mathbf{g} = \mathbf{0}$.

**Excited state landscape of azobenzene**

As our second test case, we study azobenzene. Despite being one of the most intensely studied photoswitches, azobenzene has been the subject of controversy in the literature.[92-94] The disputes over azobenzene photochemistry generally arise from the wavelength dependence, since excitation of *trans*-azobenzene to the $S_1$ ($n \rightarrow \pi^*$ character) and $S_2$ ($\pi \rightarrow \pi^*$ character) states, respectively, does not lead to the same quantum yield of *cis*-azobenzene. To properly model this behavior from first principles, it is necessary to accurately capture the potential energy landscapes across at least the $S_0$, $S_1$, and $S_2$ surfaces as well as crossings between them. Here, we aim to assess the ability of the LUSH approach to model the ground- and excited state potential energy surfaces of azobenzene by training it on a set of ~781k samples obtained with hole-hole Tamm-Dancoff approximated BHLYP-TDDFT with fractional-occupation molecular orbitals (FOMO-hh-TDA)[95]/def2-SVP[87]. This level of theory was previously used to model the photochemistry of azobenzene in excellent agreement with experiments.[94] The full data set generated here contains 780745 single point electronic structure calculations on azobenzene. The dataset is split into training and validation sets by randomly selecting 1% of the data points as the validation set yielding a train:validation split of 99:1.

In Ref. [94], it was found that the excited state decay pathways of azobenzene involve either torsion around the central double bond leading to *cis*-azobenzene or stretching/compression of it in an unreactive planar geometry. Figure 3 therefore displays absorption spectra as well as potential energy surfaces at fixed values of the $\theta_{\text{CNNC}}$ dihedral or at fixed values of the symmetric $\alpha_{\text{CNN}} = \alpha_{\text{NNC}}$ angle comparing LUSH to the FOMO-hh-TDA results from Ref. [94]. Here, it should be emphasized that the geometries in the scans are relaxed according to FOMO-hh-TDA. As seen in Figure 3A, the trained LUSH model clearly reproduces the reference FOMO-hh-TDA absorption spectra for both *trans*- and *cis*-azobenzene both in terms of relative peak positions and relative intensity. A subtle detail is that LUSH swaps the almost degenerate $S_3$ and $S_4$ states for *trans*-azobenzene (see Table S1). Since $S_3$ is dark and $S_4$ is bright, we include an additional state in the FOMO-hh-TDA spectrum for a more fair comparison of the matching states. Besides that, our trained model provides nearly identical potential energy surfaces along the $\theta_{\text{CNNC}}$ across all three states when it is relaxed according to the $S_0$ state as evidenced by Figure 3B. As the geometries are relaxed according to the $S_1$ state in Figure 3C and D, the resemblance between the LUSH and FOMO-hh-TDA results is equally good.

If the scans are relaxed according to the LUSH model, the agreement with the reference data improves for the planar angle scan (see Fig. S5). However, for the dihedral scans, the description of twisted geometries is poor. This can be rationalized by the present unbalanced training set, featuring few *cis*-like geometries (Fig. S6). The reason for this is that most of the data are generated from nonadiabatic dynamics trajectories starting from *trans*-azobenzene Wigner samples. It is consequently unsurprising that twisted geometries, including the *cis*-azobenzene minimum, are not described with the same accuracy by the model. We expect this will be significantly improved in future work through better sampling of twisted geometries and inclusion of gradient information in the training data. Nevertheless, the trained model clearly captures the same qualitative features even though it is not trained on gradients and it therefore only obtains the local curvature of the potential energy surfaces through interpolation.

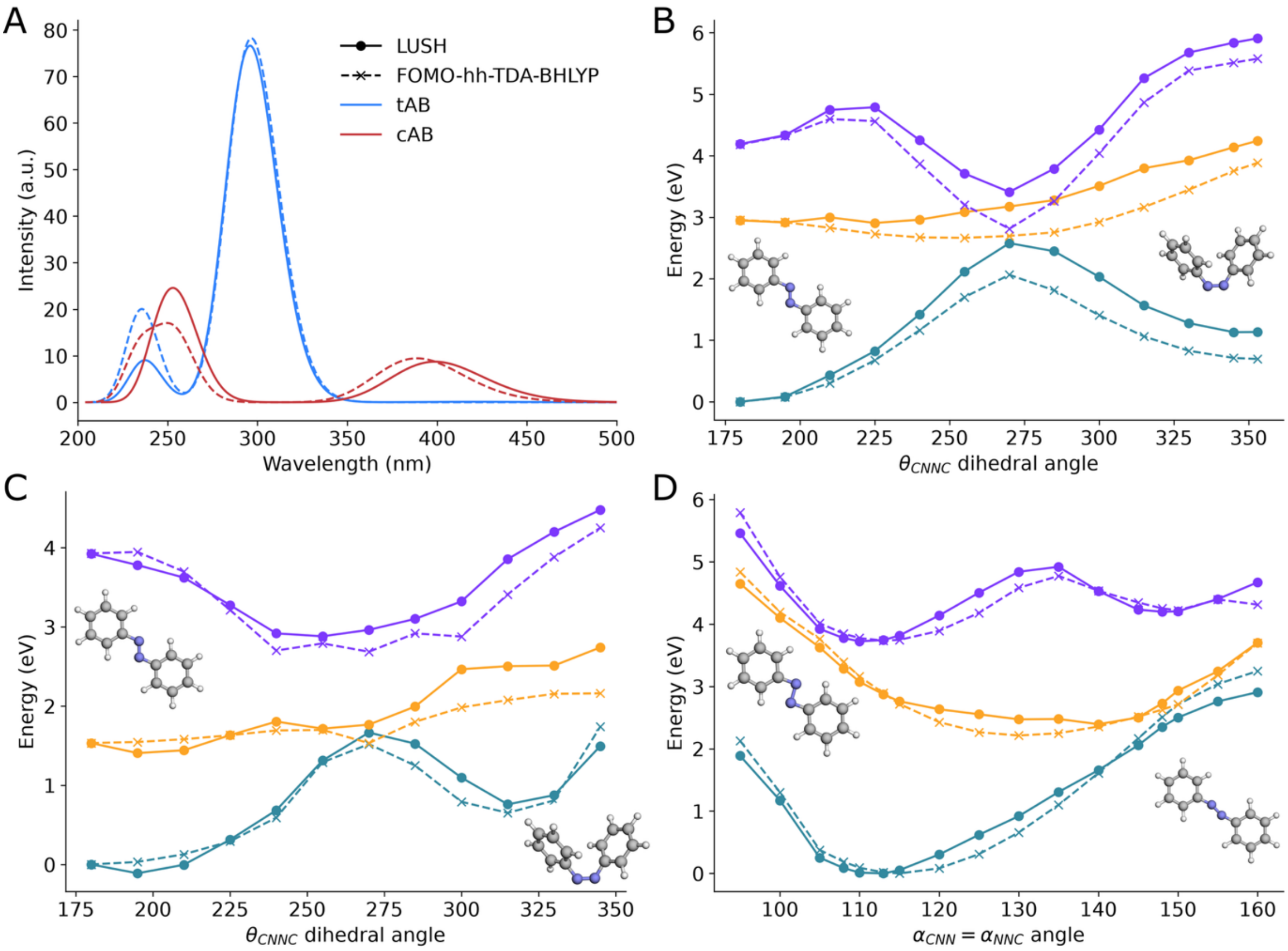

**Figure 3: Comparison of LUSH and FOMO-hh-TDA-BHLYP/def2-SVP for azobenzene** (A) Predicted absorption spectra of *trans*- and *cis*-azobenzene (tAB and cAB, respectively) generated from convolved vertical excitation energies and oscillator strengths. (B)-(C) $S_0$, $S_1$, and $S_2$ potential energy surfaces at fixed values of $\theta_{CNNC}$ relaxed according to $S_0$ and $S_1$, respectively. (D) $S_0$, $S_1$, and $S_2$ potential energy surfaces at fixed values of the symmetric $\alpha_{CNN} = \alpha_{NNC}$ angle relaxed according to $S_1$.

An important detail in the FOMO-hh-TDA potential energy surfaces in Figure 3C and D is that they contain the reactive and unreactive $S_0/S_1$ conical intersections, respectively. Meanwhile, the LUSH data has gaps at these two geometries. To assert that the trained model actually reproduces the relevant conical intersection as well as other important geometries on the $S_0$, $S_1$, and $S_2$ potential energy surfaces, we optimized all the critical point geometries given in the SI of Ref. [94] with LUSH. Figure 4 shows an overlay of the reference and LUSH geometries, the root mean square deviation (RMSD) between corresponding atoms in cartesian space after alignment, and the error in the relative energy compared to *trans*-azobenzene on the $S_0$ state ($\Delta\Delta E$).

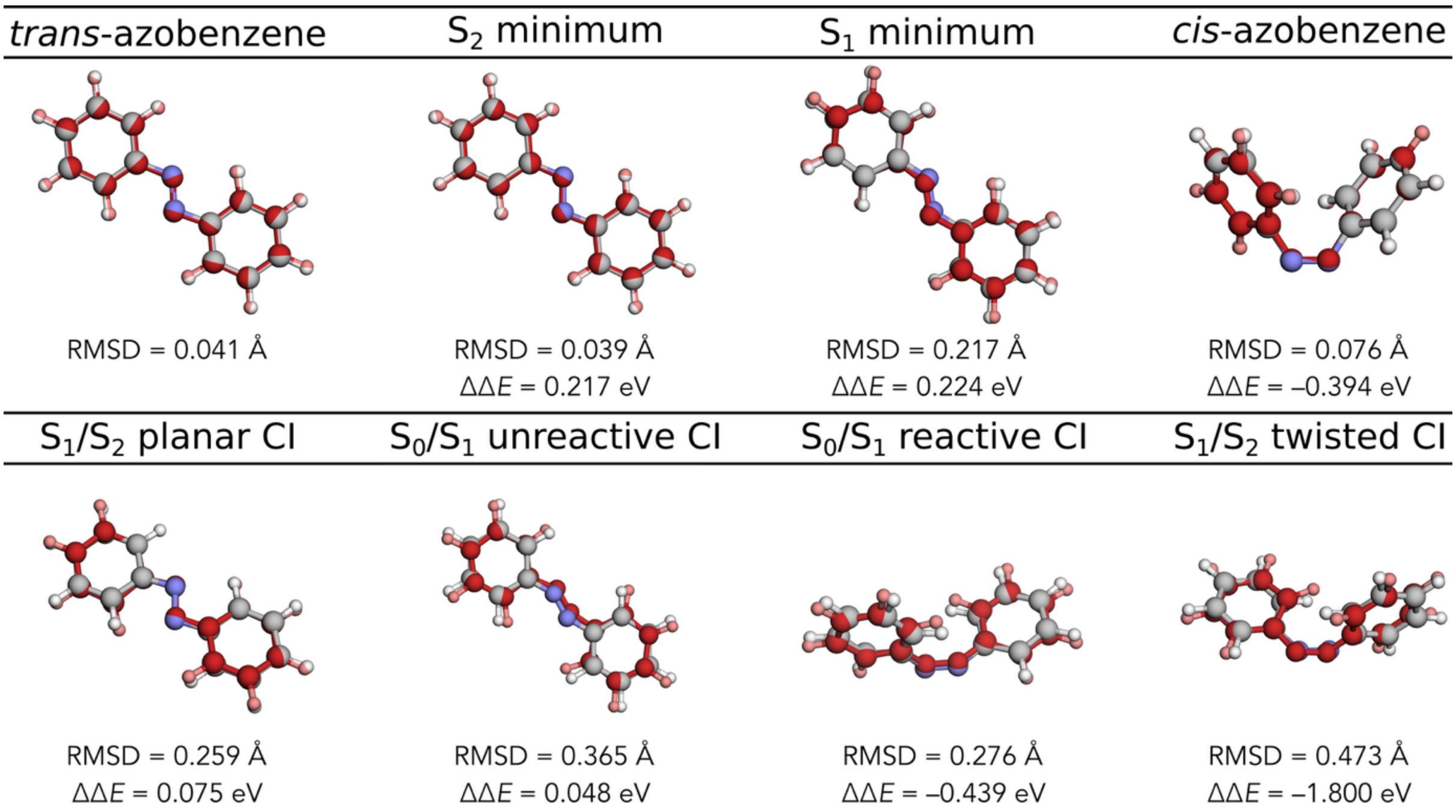

**Figure 4: Performance of LUSH model for critical geometries in excited state manifold of azobenzene.** Superimposed critical geometries as predicted by FOMO-hh-TDA-BHLYP/def2-SVP (grey/blue) and LUSH (red), respectively, along with the root mean square deviation (RMSD) between corresponding atoms and the error in the relative energy compared to *trans*-azobenzene on the $S_0$ state ($\Delta\Delta E$).

As seen from Figure 4, the LUSH optimized geometries are very similar to the reference FOMO-hh-TDA geometries. For the minima, the RMSD values are small and so are the relative energy errors. It is consequently evident that LUSH accurately reproduces the different mimima across the $S_0$, $S_1$, and $S_2$ states. Meanwhile, the LUSH MECIs deviate slightly more from the reference geometries with RMSD values up to 0.473 Å. While the energetics for the planar $S_1/S_2$ CIs are in excellent agreement with the reference, $\Delta\Delta E$ can reach values of -1.8 eV for twisted geometries. This discrepancy is not surprising considering that MECI search is a much harder optimization problem and even small differences in the potential energy surfaces may lead to large changes in the CI seam space. Even among different flavors of TDDFT, differences of up to 0.3 eV are expected. Even more importantly, the training data included a limited amount of low-energy *cis*-like configurations (Fig S6A). As a result, LUSH understabilizes twisted geometries (*cf.* Figure 3 above), including the MECIs. With this in mind, the MECI structures and energetics predicted by LUSH are in overall good agreement with FOMO-hh-TDA, indicating it is able to correctly capture the relevant CIs.

## Conclusion and Outlook

The present work introduces the Latent Unified Smooth Hamiltonian (LUSH) neural network framework for the prediction of electronic ground and excited states of molecular systems by constructing a latent electronic-state Hamiltonian. Diagonalization of the predicted

Hamiltonian matrix yields adiabatic electronic structure properties including state energies and their intersections. Beyond energetics and derivative properties such as atomic forces and NACs, LUSH provides a unified framework for the prediction of general latent operators such as (transition) dipoles and their evaluation in the adiabatic frame (see Figure 1). The presented methodology thus expands upon the philosophy of physically interpretable latent variables such as latent effective charges for explicit long-range (electrostatic) interactions introduced in previous works.[96-100]

The LUSH methodology rests on four key components. First, we employ a cross-attention mechanism with a fixed number of trainable latents, which allows us to encode arbitrary-size molecules into a fixed-size set of tokens (feature vectors). These tokens can be viewed as raw features associated with the individual electronic states, further refined through self attention. Second, the resulting single-state representation is transformed into an explicit pairwise representation, which is subsequently refined by the Pairformer block efficiently incorporating global communication before Hamiltonian generation and diagonalization. Third, a differentiable eigensolver applicable to systems with degenerate states assures stable and finite gradients near conical intersections. Lastly, the weighted inclusion of state and transition properties based on their energy gap to the spectrum boundary allows us to limit the approach to a fixed-size low-dimensional Hamiltonian submatrix without contaminating low-lying states when encountering state degeneracies at the spectrum boundary.

Using LUSH, we trained a series of models that confirm its applicability to accurately predict both ground and excited electronic states as well as the conical intersections between them. For the QM9 data set, LUSH provides a model that readily achieves chemical accuracy for ground states. In addition to that, we used the LUSH architecture to train models that reproduce the excited state energy landscapes of thymine and azobenzene, both of which are prototype systems in photochemistry. For these systems, our results demonstrate that LUSH provides a uniform description of ground and excited electronic states and transitions between them. The models reproduce the important features of the potential energy surfaces involved in their photochemistry as well as their absorption spectra. Importantly, the LUSH models include a correct description of conical intersections between the electronic states including the topography of NACs. Altogether, our test cases show that LUSH can provide accuracy on par with quantum chemical reference data, but at a fraction of the computational cost.

Leveraging LUSH as an efficient tool for excited state calculations including NACs, future work will explore its capability to push the boundary of nonadiabatic dynamics simulations to extended time and length scales. In this vein, we will expand training to include energy gradients and (phaseless) NACs. Based on the already very good description of critical geometries by the current models using a gradient-free loss, we expect this extended loss definition to further boost the performance. Future work will also explore possible ML/MM setups, coupling LUSH to an even more efficient molecular mechanics description of the environment to further extend tractable system sizes towards biomolecular complexes. Another interesting question is in how far the imposed mathematical structure in LUSH can “correct” spurious reference data. For example, whether sufficient information on the electronic state characters enables LUSH to reproduce conical intersections that are absent in the reference data (*e.g.*, $S_0/S_1$ intersections missed by TDDFT[101]). Encoding the general mathematical structure of the multi-state electronic structure problem rather than imposing restrictive constraints, we believe LUSH offers a good balance between inductive bias from quantum chemistry and the representation learning power of deep networks.

## Methods and Computational Details

### *Additional details on LUSH Pairformer*

To construct the initial pair representation, $\mathcal{Z} \in \mathbb{R}^{m\times m\times d}$, we first perform the "pair lift" operation on $\mathbf{S}'$ defined in Equation (1). This produces $\mathcal{R} \in \mathbb{R}^{m\times m\times 3d}$. We then use a single layer perceptron to project the hidden dimension of $\mathcal{R}$ $(3d)$ back down to $d$, followed by application of a single resnet to yield the initial state-pair representation $\mathcal{Z} \in \mathbb{R}^{m\times m\times d}$. $\mathcal{Z}$ is then passed into an in-house implementation of the pairformer block introduced in AlphaFold3 (AF3).[63] Briefly, the AF3 pairformer performs axial (row-wise and column-wise) triangle multiplication and attention updates on $\mathcal{Z}$ to yield $\mathcal{Z}' \in \mathbb{R}^{m\times m\times d}$, with overall computational scaling $\mathcal{O}(m^3)$. A detailed description of the AF3 pairformer block goes beyond scope of this publication, and the interested reader is referred to the Supporting Information of Ref. [63]. For most photochemistry applications (such as the ones examined in this study) the number of treated electronic states can remain $m < 10$, and thus the cost of triangle attention in LUSH remains very modest.

Once refined by the pair attention stack, parallel heads project the state-pair representation $\mathcal{Z}'$ down to the latent operator representations $\mathbf{O} \in \mathbb{R}^{m\times m\times d_o}$, where $d_o$ is the dimensionality of each specific operator $\hat{O}$. In our study we have two operator representations: the Hamiltonian matrix $\mathbf{H}$ with $d_o = 1$, and the transition dipole tensor $\mu$ with $d_o = 3$, representing the $x, y$ and $z$ directions. Thus, to produce $\mathbf{H}$ we use a linear layer (no activation) with output dimension one to project $\mathcal{Z}' \in \mathbb{R}^{m\times m\times d} \mapsto \mathbf{H} \in \mathbb{R}^{m\times m}$, and we use a resnet followed by a linear layer (no activation) with output dimension three to transform $\mathcal{Z}' \in \mathbb{R}^{m\times m\times d} \mapsto \mu \in \mathbb{R}^{m\times m\times 3}$.

To remove the requirement for the pair attention stack to perform the absolute scaling of the state energies, we subtract the trace of $\mathbf{H}$ from its diagonal entries immediately after generating it. Then, we project the refined single-state representation $\mathbf{S}' \in \mathbb{R}^{m\times d}$ to a vector $\mathbf{c} \in \mathbb{R}^m$ using an "offset head" (resnet followed by a linear projection with output dimension of one). Finally, we add the mean value of $\mathbf{c}$ to the diagonal elements of the now traceless $\mathbf{H}$. This reintroduces a learned geometry-dependent absolute shift to the spectrum of the Hamiltonian, improving the description of *intra*-state relative energies and allowing the pairformer to focus on the excitation energies.

### *Details of loss definition*

The first contribution to the loss function given in Eq. (2) denotes the mean-square error in atomization energies across all states and all training samples,

$$L_E = \frac{1}{n_{\text{ref}}} \sum_{I=1}^{n_{\text{ref}}} \frac{1}{m} \sum_{i=0}^{m-1} w_i \left(E_i^{(I)} - \bar{E}_i^{(I)}\right)^2 \tag{4}$$

with $\bar{E}$ denoting the corresponding *ab initio* reference energy and $w_i$ being the state weight according to Eq. (3). State-resolved reference energies are thereby calculated as $\bar{E}_i = \varepsilon_i - \sum_A E_{\text{iso}}^{(A)}$, where $\varepsilon_i$ is the total electronic energy of state $i$ as predicted by the reference method. For the QM9 database, we use isolated-atom energies, $E_{\text{iso}}$, from Hartree-Fock. For the QeMFi dataset, we obtain effective single-atom energies by fitting the sum of fixed atom energies to best reproduce the ground-state reference energy across the whole training set. For azobenzene, we calculate $E_{\text{iso}}$ for the corresponding atom *in vacuo* using BHLYP/def2-SVP. For further details on generation of atomic reference energies, see supporting information. The excitation energy contribution, $L_\Delta$, and the oscillator strength loss, $L_{\text{osc}}$, are calculated as the mean smooth-L1 (scaled Huber) loss,

$$L_E = \frac{1}{n_{\text{ref}}} \sum_{I=1}^{n_{\text{ref}}} \frac{2}{m(m-1)} \sum_{i,j>i} w_i w_j \cdot l_{ij}^{(I)}$$
$$\text{with } l_{ij}^{(I)} = \begin{cases} \frac{1}{2\beta}\left(y_{ij}^{(I)} - \bar{y}_{ij}^{(I)}\right) & \text{if } \left|y_{ij}^{(I)} - \bar{y}_{ij}^{(I)}\right| < \beta \\ \left|y_{ij}^{(I)} - \bar{y}_{ij}^{(I)}\right| & \text{else} \end{cases} \tag{5}$$

where $y_{ij}$ represents either $\Delta_{ij}$ for excitation energies or $f_{ij}$ for the oscillator strength loss. We use $\beta = 10^{-3}$ eV and $\beta = 10^{-3}$ a.u. for the excitation energy and oscillator strength losses, respectively.

### *Geometry optimization and conical intersection search*

All azobenzene geometry optimizations in this work were performed using a Levenberg-Marquardt-type quasi-Newton scheme with delocalized internal coordinates[102] as implemented in the geomeTRIC software package.[103] The default convergence criteria are used for geometry optimizations, however, for MECI searches and relaxed surface scans the gradient tolerance was loosened to 1.7e-3 for the root mean square of the gradient and 1.0e-3 for the maximum gradient component. The search for azobenzene MECI structures followed a penalty function-based approach. MECIs are thereby identified by a vanishing energy gap $\Delta_{ij}$ and simultaneously minimal average state energy $\left(E_i + E_j\right)/2$. As detailed in Ref. [19], for a two-state intersection this can be phrased as the minimization of the objective/cost function,

$$c(\mathbf{R}; i, j) = \frac{1}{2}\left(E_i(\mathbf{R}) + E_j(\mathbf{R})\right) + \sigma \cdot \frac{\Delta_{ij}^2(\mathbf{R})}{\Delta_{ij}(\mathbf{R}) + \alpha} \tag{6}$$

where $\sigma$ determines the weight of the gap penalty and $\alpha$ represents a regularization parameter. For the present work, we chose $\sigma = 10$ and $\alpha = 10^{-4}$, which provided a good balance of regularization and gap closure. The MECI search was deemed converged according to the above outlined criteria.

### *Azobenzene reference calculations*

The training geometries for azobenzene are sampled from 423 K harmonic Wigner distributions around the *cis*- and *trans*-azobenzene minima as well as non-adiabatic molecular dynamics of azobenzene including the $S_0$, $S_1$, and $S_2$ states following excitation of *trans*-azobenzene to either the $S_1$ state or the $S_2$ state. Dynamics simulations starting from elevated-temperature initial conditions thereby sample a broad range of the excited state landscape. The electronic structure was modeled using FOMO-hh-TDA-BHLYP[95]/def2-SVP[87], which was previously demonstrated to achieve excellent accuracy for the UV absorption and emission spectra of azobenzene.[94] The *ab initio* multiple spawning method[104,105] was used to simulate the nonadiabatic dynamics. All electronic structure calculations for azobenzene were performed with the TeraChem software package.[88,89] Based on the hh-TDA reference data for states $S_0$ through $S_4$, we constructed a five-state LUSH model with $S_3$ and $S_4$ as buffer states, assuring a framework able to correctly reproduce the states $S_0$, $S_1$, and $S_2$ studied in this work.

## Supplementary information

Additional details and analysis on QeMFi and azobenzene databases, calculation of atomic reference energies, and secondary results including Figs. S1 to S6 and Tab. S1.

## Acknowledgments

This work was supported by the AMOS program of the U. S. Department of Energy, Office of Science, Basic Energy Sciences, Chemical Sciences and Biosciences Division. DJ acknowledges financial and professional support from the Stanford Energy Postdoctoral Fellowship and the Stanford Precourt Institute for Energy. MS acknowledges financial support from the Deutsche Forschungsgemeinschaft (DFG, German Research Foundation) -- 534068594. AEHB acknowledges financial support from the Novo Nordisk Foundation under grant reference number NNF24OC0089345. OJF is a Department of Energy Computational Science Graduate Fellow (Award Number DE-SC0023112), supported by the U.S. Department of Energy, Office of Science, Office of Advanced Scientific Computing Research.

## Competing interests

The authors declare the following competing financial interest(s): TJM is a cofounder of PetaChem, LLC.

## Data and code availability

The source code for training and inference with LUSH will be made publicly available upon acceptance of this work in a peer-reviewed journal.

## Author contributions

DJ: Development of original idea; design of network architecture (primary); implementation (primary); data acquisition; analysis; writing (initial draft and editing).

MS: Development of original idea; design of network architecture (support); implementation (support); generation of reference data (support); analysis; writing (initial draft and editing).

AEHB: Data acquisition; conception of test cases; analysis; writing (initial draft and editing).

OJF: Generation of reference data (primary); conception of test cases; writing (editing).

TJM: Acquisition of funding; writing (editing); supervision.

# Supplementary Materials for
# Latent unified smooth Hamiltonians for excited state chemistry

David Juergens, Martin Stöhr, Andreas E. Hillers-Bendtsen,
O. Jonathan Fajen, Todd J. Martínez*
*Corresponding author. Email: todd.martinez@stanford.edu

## Supplementary Figures

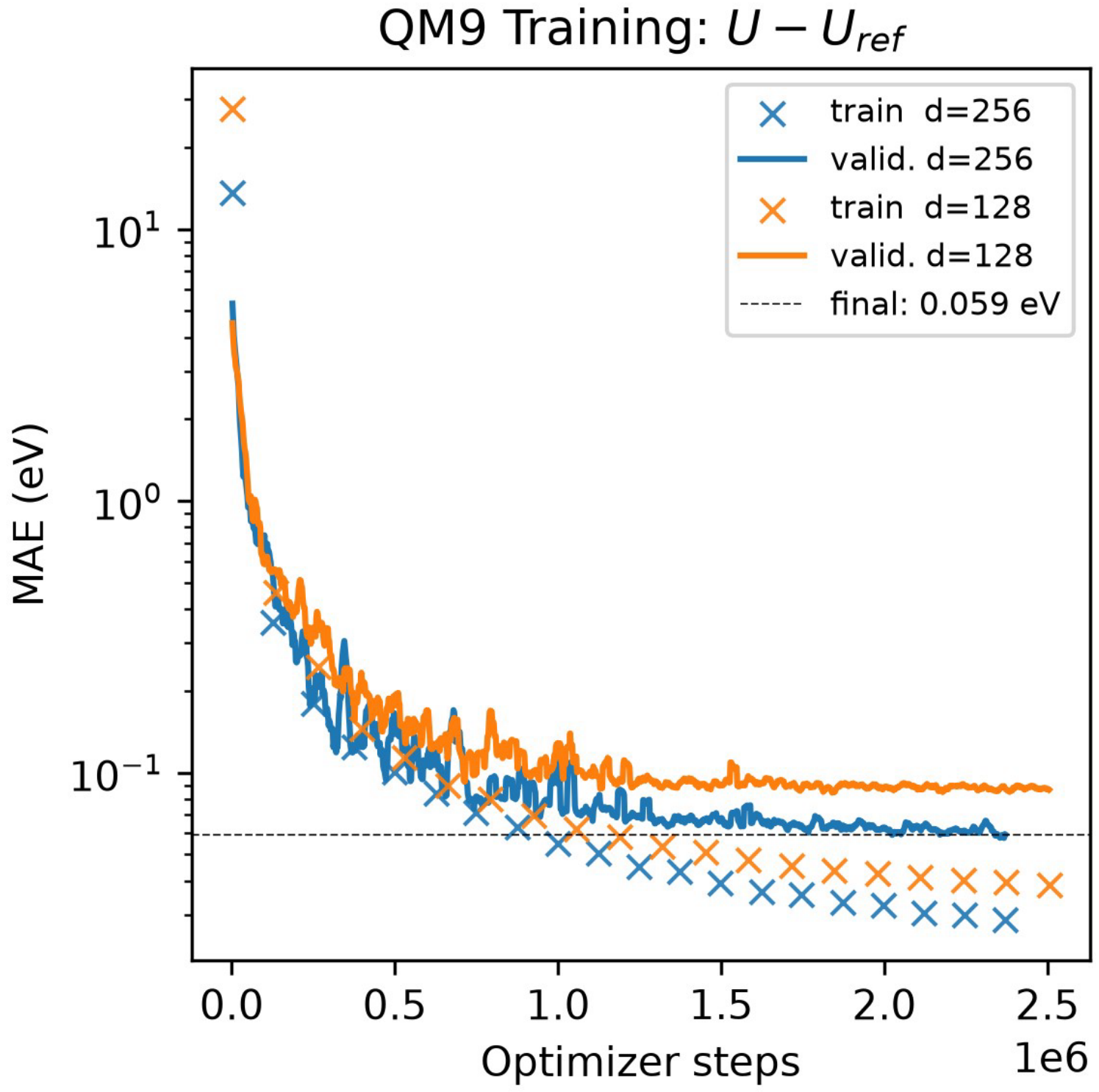

**Fig. S1**: **Training LUSH on the QM9 ground state DFT dataset.** We trained two LUSH models on the QM9 total potential energy ("$U$") prediction task. To normalize the total energies in the dataset, we trained on the quantity $U - U_{ref}$, where $U_{ref}$ is the sum of the isolated atomic energies in the molecule (a constant per molecular formula). The models were of identical hyperparameters aside from their hidden dimensions (256, and 128). As seen by the final validation error observed, the ~7.05 million parameter (d=256) model approaches 0.059 eV (~1.36 kcal/mol) in validation set $U - U_{ref}$ mean absolute error.

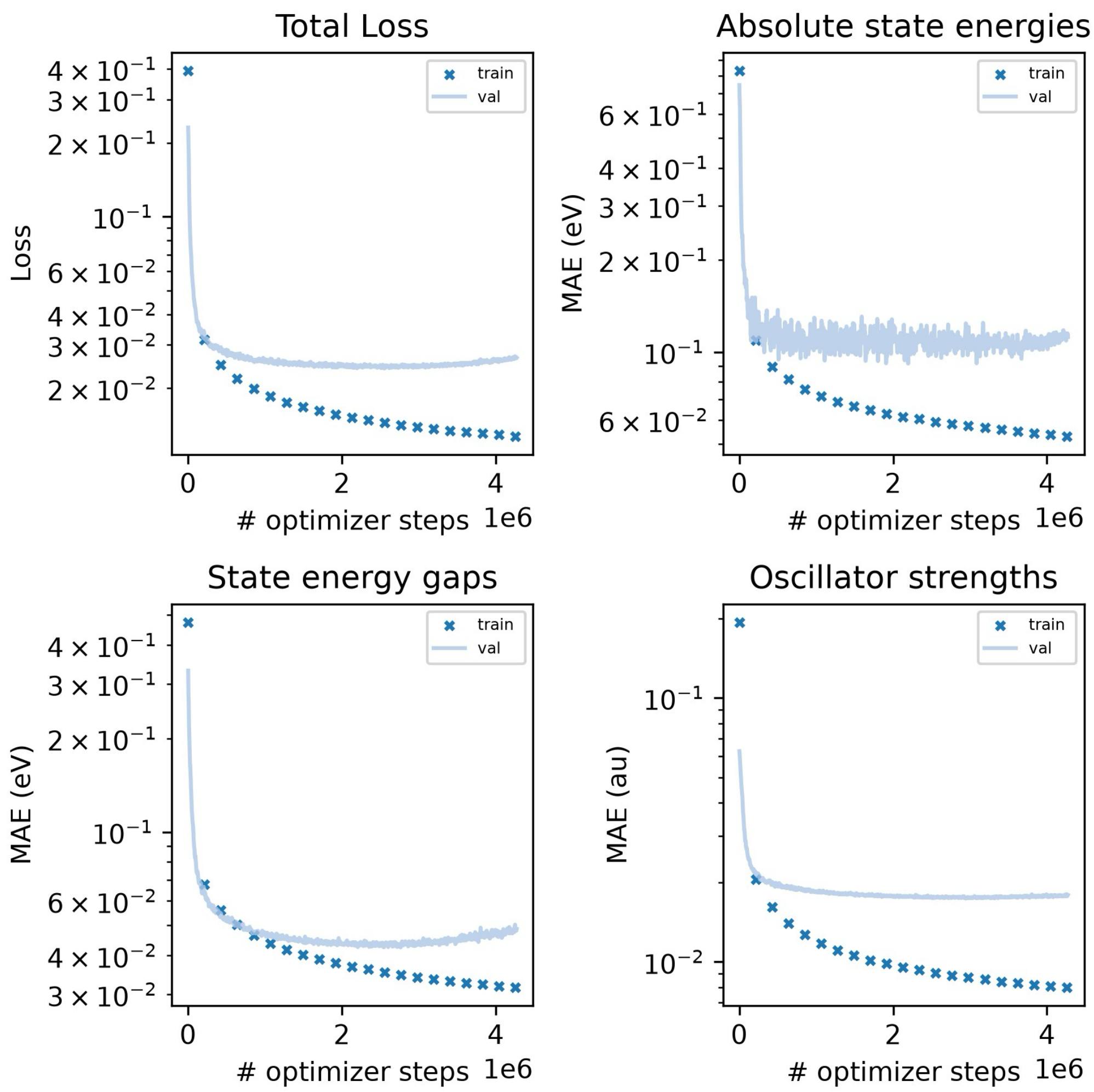

**Fig. S2**: **Training LUSH on the QeMFi dataset.** Performance on total loss, state energies, state energy gaps, and oscillator strengths while training LUSH on the QeMFi def2-TZVP dataset (135,000 samples from all nine molecules). Train and test datasets were constructed through a 90/10 random split of the full set.

**Table S1**: Vertical excitation energies (*E*, in eV) and oscillator strengths ( *f* ) for *trans*- and *cis*-azobenzene.

| | | LUSH | | FOMO-hh-TDA-BHLYP/def2-SVP | |
|---|---|---|---|---|---|
| **Isomer** | **State** | *E* (eV) | *f* | *E* (eV) | *f* |
| *trans*-Azobenzene | $S_1$ | 2.95 | 0.002 | 2.95 | 0.000 |
| | $S_2$ | 4.19 | 1.422 | 4.18 | 1.451 |
| | $S_3$ | 5.22 | 0.168 | 5.24 | $4.1 \times 10^{-7}$ |
| | $S_4$ | - | - | 5.26 | 0.372 |
| *cis*-Azobenzene | $S_1$ | 3.11 | 0.162 | 3.19 | 0.175 |
| | $S_2$ | 4.77 | 0.247 | 4.88 | 0.270 |
| | $S_3$ | 5.00 | 0.289 | 5.28 | 0.224 |
| | $S_4$ | - | - | 5.53 | 0.422 |

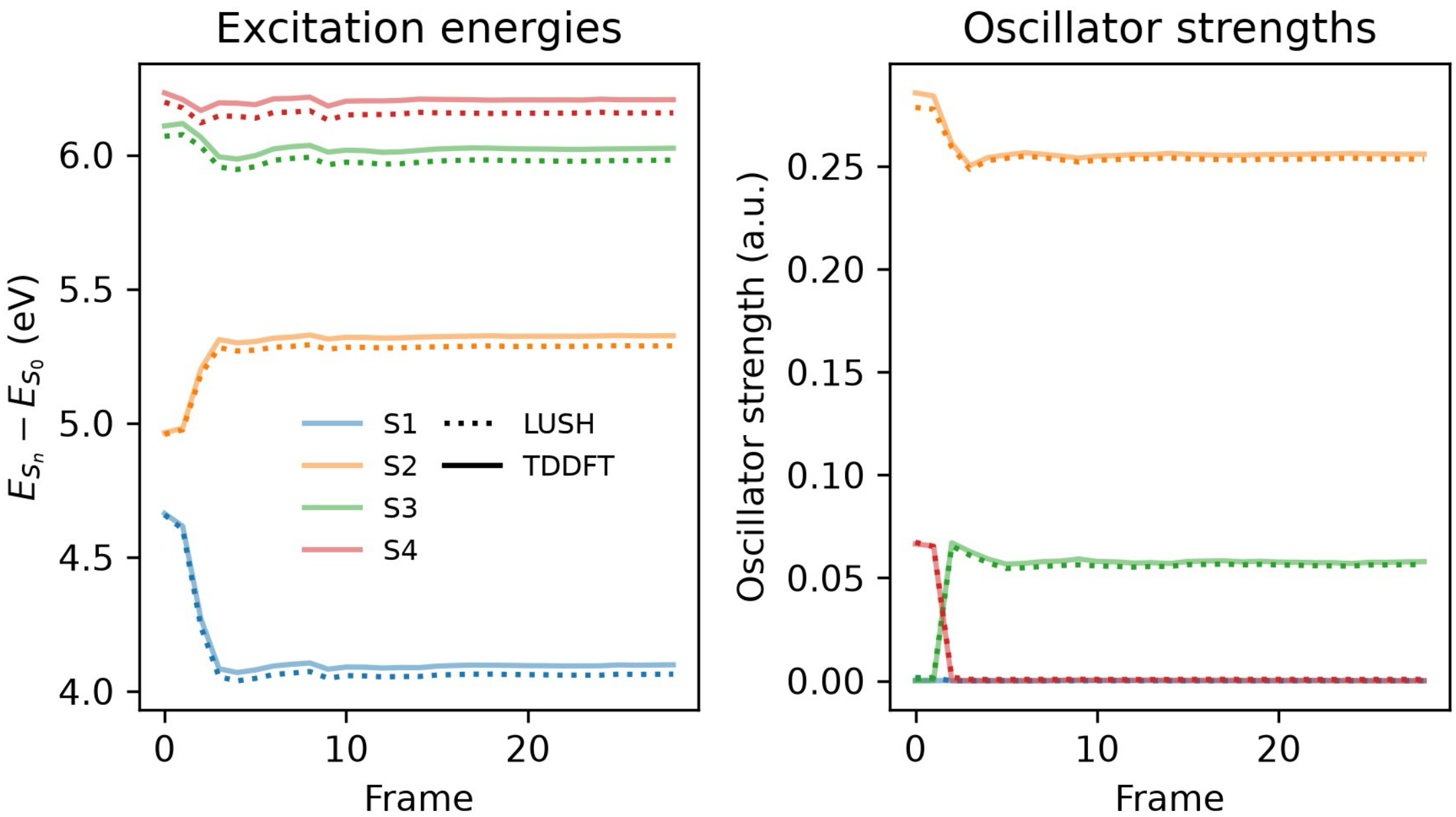

**Fig. S3**: **Tracking electronic state crossings with LUSH.** Staring from a higher-level reference thymine $S_1/S_2$ MECI geometry, optimization on the TDDFT $S_1$ surface was performed. Predictions of excitation energies (left) and oscillator strengths (right) for all geometries from this trajectory were then made with a LUSH model pretrained on the QeMFi def2-TZVP set. Although the geometries from the optimization trajectory were not in the training set, the values predicted by LUSH track closely to TDDFT. Furthermore, LUSH correctly tracks an $S_3/S_4$ crossing is observed from frame index 1 to 2, as indicated by the switching of predicted $S_3/S_4$ oscillator strengths to match the reference. Note that plots for $S_5$ were omitted for clarity because the $S_5$ oscillator strengths obscure viewing of the $S_3$ oscillator strengths, the $S_6$ plots were omitted because it corresponds to the second to last $(m-1)$ state produced by the model which has low weight in the loss function in regions of small $m$ to $m+1$ energy gap, and the $S_7$ plots were omitted because it corresponds to the untrained buffer state $(m)$.

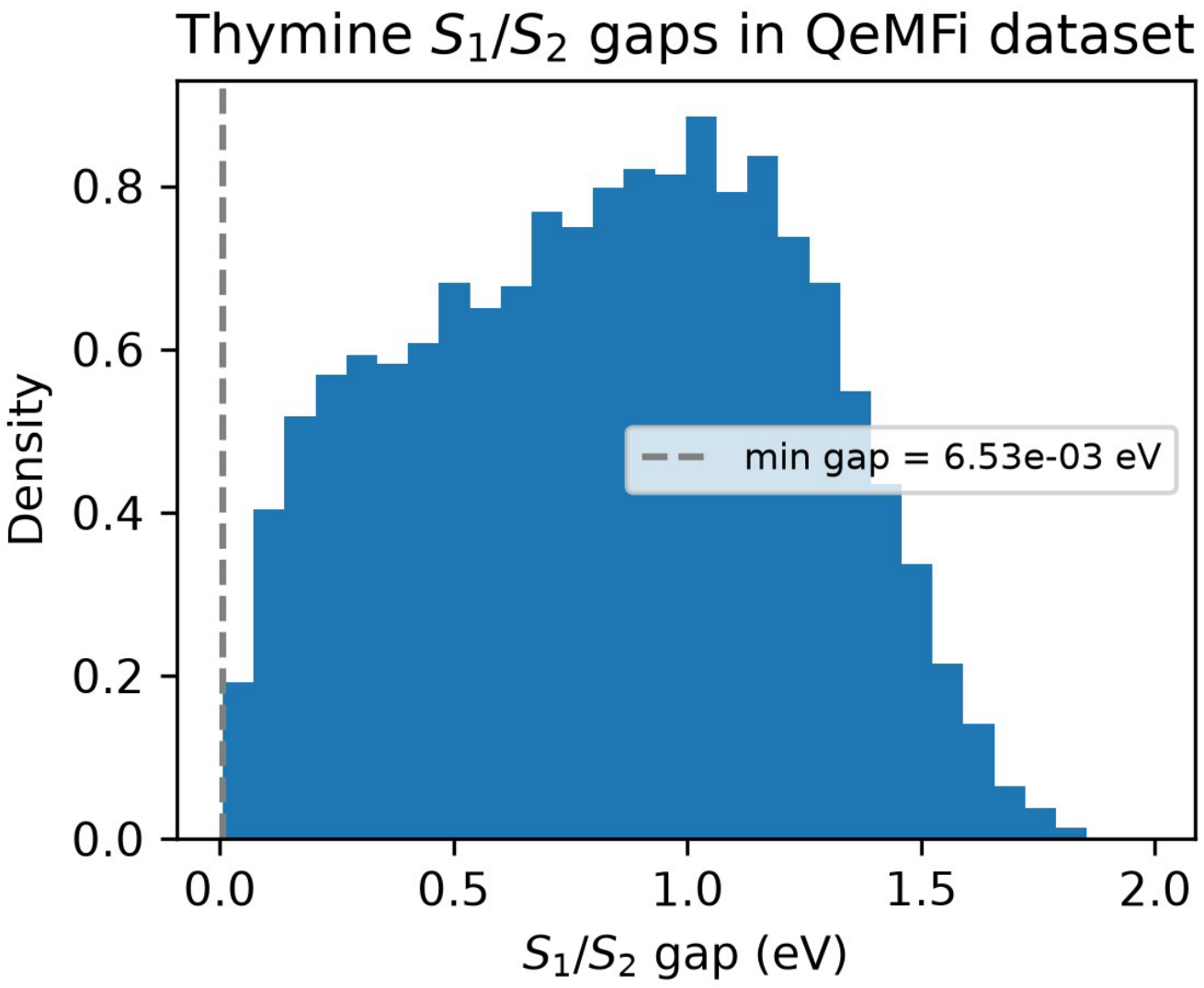

**Fig. S4**: **QeMFi Thymine def2/TZVP dataset.** A histogram of $S_1/S_2$ energy gaps for all 15,000 def2-TZP calculations on thymine in the QeMFi dataset. The smallest energy gap observed in the entire set is 6.53e-03 eV.

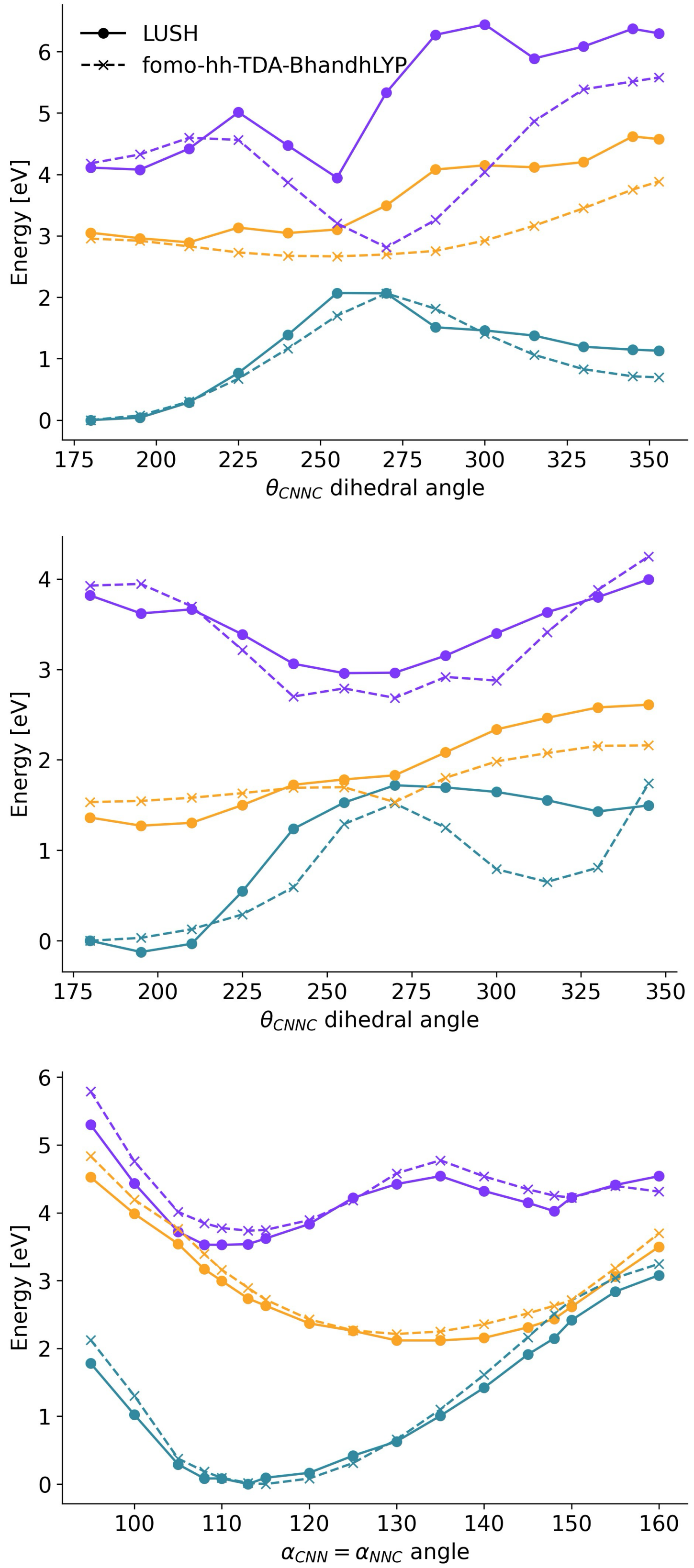

**Fig. S5**: **Comparison of LUSH and FOMO-hh-TDA-BHLYP/def2-SVP for azobenzene with reoptimized LUSH scans.** (Top and Middle) $S_0$, $S_1$, and $S_2$ potential energy surfaces at fixed values of the $\theta_{\mathrm{CNNC}}$ relaxed according to $S_0$ and $S_1$, respectively. (Bottom) $S_0$, $S_1$, and $S_2$ potential energy surfaces at fixed values of the symmetric $\alpha_{\mathrm{CNN}} = \alpha_{\mathrm{NNC}}$ angle relaxed according to $S_1$.

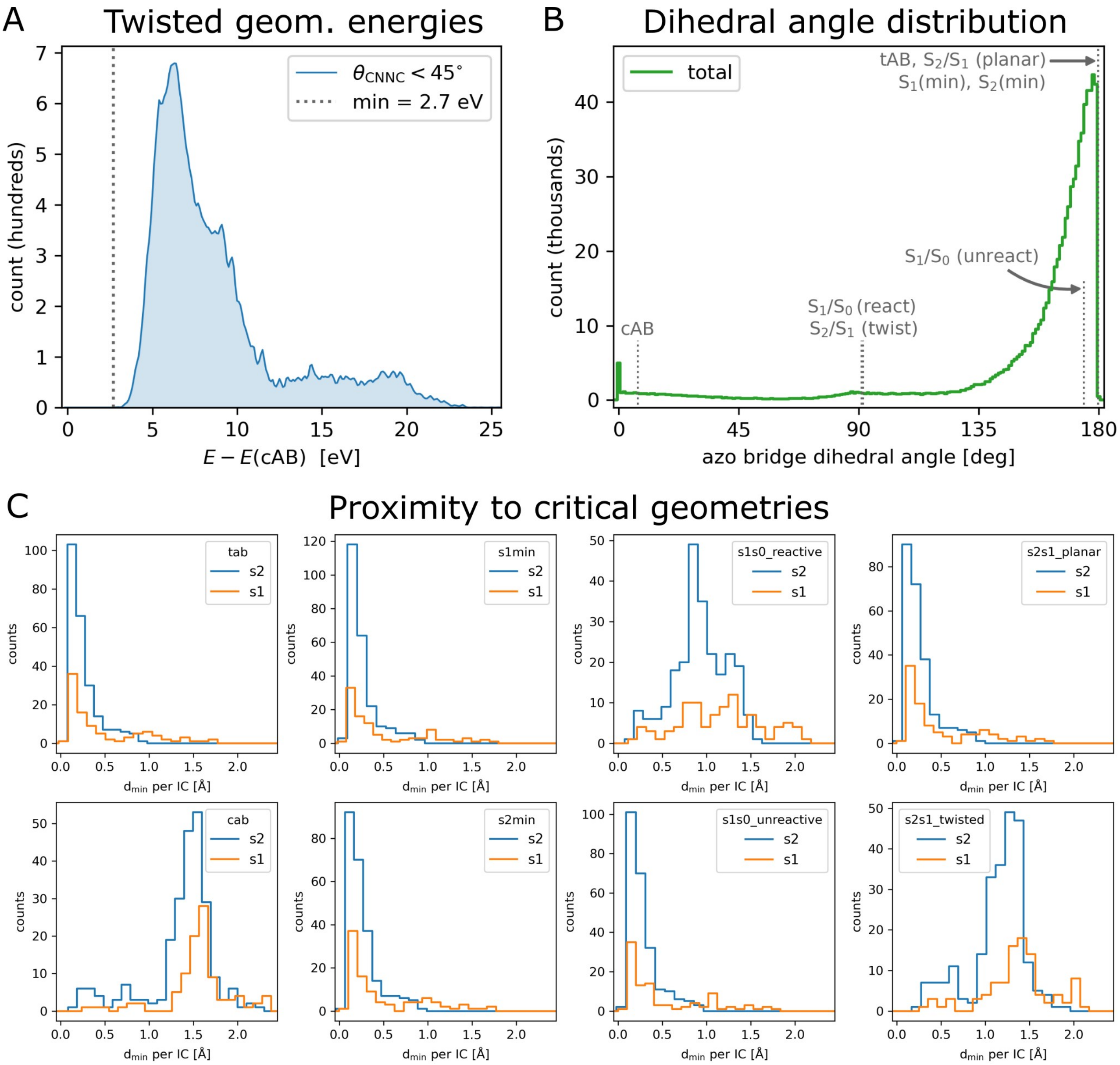

**Fig. S6**: **Proximity of ~781k azobenzene training samples to critical geometries.** The training data used to train LUSH on azobenzene is derived from *ab initio* multiple spawning molecular dynamics (MD) trajectories from randomly sampled initial conditions (ICs). (A) Distribution of sample energies relative to *cis*-azobenzene (cAB) for all geometries containing CNNC dihedral angle <45°. The vast majority of these samples lie very high in energy above cAB, with the closest sample at 2.7 eV above cAB. (B) Distribution of CNNC dihedral angle within the training set, with the dihedral angles of critical azobenzene geometries labeled. (C) Distribution of the closest (minimum cartesian coordinate RMSD after a symmetry-aware Kabsch alignment) geometry observed across the unique MD trajectories (“per IC”) to each of the critical azobenzene geometries, colored by the electronic state the IC began on. Note that zero geometries within 0.2Å are observed across all data for the cAB, $S_1/S_0$ reactive conical intersection (CI), and $S_2/S_1$ twisted CI.

**Computation of atomic reference energies**

In the current work, there is a unique set of atomic reference energies used for each of the three datasets discussed in the paper (QM9, QeMFi, Azobenzene) as listed in Tab S2. For the QM9 dataset, reference energies for isolated atoms were computed at the Hartree-Fock/cc-pv5z level of theory. For the QeMFi dataset, reference energies are computed by fitting a linear model to map the chemical formula of a molecule to its total potential energy. Specifically, we solve the least squares minimization problem

$$\hat{\epsilon} = \underset{\epsilon}{\operatorname{argmin}} \|A\epsilon - \mathbf{b}\|_2^2 \tag{S1}$$

where $K$ is the number of unique atom types being fit, $N$ is the number of training geometries, and $\mathbf{b} \in \mathbb{R}^N$ is the list of target ground state energies for each geometry such that the $i$'th element of $\mathbf{b}$ is the $S_0$ energy of the $i$'th molecule in the dataset, and the composition matrix $A \in \mathbb{R}^{N\times K}$ is

$$A_{ij} = \sum_{a\in\mathcal{A}_i} \mathbb{I}[Z_a = Z_j] \tag{S2}$$

where $\mathcal{A}_i$ is the set of atoms in the molecule associated with geometry $i$, $\mathbb{I}$ is the indicator function, $Z_a$ is the atomic number of atom $a$, and $Z_j$ is the atomic number corresponding to the element column $j$. Thus, $A_{ij}$ is simply the number of times at atom with atomic number $Z_j$ occurs in sample $i$. The minimizer $\hat{\epsilon}$ is the final set of reference atomic energies. For the azobenzene dataset, atomic reference energies were computed on single atoms at the ROKS BHLYP/def2-SVP level of theory.

**Table S2**: **Atomic reference energies used for different datasets.** QM9: Single atom ground state energies ($E_{iso}$) computed at the Hartree-Fock/cc-pv5z level of theory. QeMFi: $E_{iso}$ from linear fit. azobenzene: $E_{iso}$ from BHLYP/def2-SVP calculations.

| | $E_{iso}$ [Hartree] | | |
|---|---|---|---|
| element | QM9 | QeMFi | azobenzene |
| H | -0.499278403420 | -0.598 | 0.4978398270 |
| He | -2.855160477243 | — | — |
| Li | -7.432420527596 | — | — |
| Be | 14.572337630953 | — | — |
| B | 24.529961624371 | — | — |
| C | 37.686544437314 | -38.044 | 37.7326788080 |
| N | 54.391114562200 | -54.695 | 54.4145730772 |
| O | 74.792166058284 | -75.129 | 74.8764634879 |
| F | 99.375240303129 | — | — |
| Ne | 128.488775551741 | — | — |
| Na | 161.853056693518 | — | — |
| Mg | 199.608297028624 | — | — |
| Al | 241.873509588698 | — | — |
| Si | 288.850076566695 | — | — |
| P | 340.709048059139 | — | — |
| S | 397.496801538240 | — | — |
| Cl | 459.471143486021 | — | — |
| Ar | 526.799865309746 | — | — |